\documentclass{article}

\usepackage{microtype}
\usepackage{graphicx}
\usepackage{booktabs} % for professional tables

\usepackage{hyperref}
\usepackage{amssymb}
\usepackage{amsmath}
\usepackage{tikz}
\usetikzlibrary{arrows.meta,positioning,shapes.geometric,fit,calc,backgrounds,decorations.pathreplacing}
\usepackage[accepted]{icml2018}

\icmltitlerunning{Offline RL for Adaptive Policy Retrieval in Prior Authorization}

\begin{document}

\twocolumn[
\icmltitle{Offline RL for Adaptive Policy Retrieval in Prior Authorization}

% It is OKAY to include author information, even for blind
% submissions: the style file will automatically remove it for you
% unless you've provided the [accepted] option to the icml2018
% package.

% List of affiliations: The first argument should be a (short)
% identifier you will use later to specify author affiliations
% Academic affiliations should list Department, University, City, Region, Country
% Industry affiliations should list Company, City, Region, Country

% You can specify symbols, otherwise they are numbered in order.
% Ideally, you should not use this facility. Affiliations will be numbered
% in order of appearance and this is the preferred way.
\icmlsetsymbol{equal}{*}

\begin{icmlauthorlist}
\icmlauthor{Ruslan Sharifullin}{su}
\icmlauthor{Maxim Gorshkov}{su}
\icmlauthor{Hannah Clay}{su}

\end{icmlauthorlist}

\icmlaffiliation{su}{Stanford University}

\icmlcorrespondingauthor{Ruslan Sharifullin}{rshar@stanford.edu}
\icmlcorrespondingauthor{Maxim Gorshkov}{maximgor@stanford.edu}
\icmlcorrespondingauthor{Hannah Clay}{hclay116@stanford.edu}

% You may provide any keywords that you
% find helpful for describing your paper; these are used to populate
% the "keywords" metadata in the PDF but will not be shown in the document
\icmlkeywords{Reinforcement Learning}

\vskip 0.3in
]

% this must go after the closing bracket ] following \twocolumn[ ...

% This command actually creates the footnote in the first column
% listing the affiliations and the copyright notice.
% The command takes one argument, which is text to display at the start of the footnote.
% The \icmlEqualContribution command is standard text for equal contribution.
% Remove it (just {}) if you do not need this facility.

\printAffiliationsAndNotice{}  % leave blank if no need to mention equal contribution
% \printAffiliationsAndNotice{\icmlEqualContribution} % otherwise use the standard text.

\begin{abstract}
Prior authorization (PA) requires interpretation of complex and fragmented coverage policies, yet existing retrieval-augmented systems rely on static top-$K$ strategies with fixed numbers of retrieved sections. Such fixed retrieval can be inefficient and gather irrelevant or insufficient information. We model policy retrieval for PA as a sequential decision-making problem, formulating adaptive retrieval as a Markov Decision Process (MDP). In our system, an agent iteratively selects policy chunks from a top-$K$ candidate set or chooses to stop and issue a decision. The reward balances decision correctness against retrieval cost, capturing the trade-off between accuracy and efficiency. We train policies using Conservative Q-Learning (CQL), Implicit Q-Learning (IQL), and Direct Preference Optimization (DPO) in an offline RL setting on logged trajectories generated from baseline retrieval strategies over synthetic PA requests derived from publicly available CMS coverage data. On a corpus of 186 policy chunks spanning 10 CMS procedures, CQL achieves 92\% decision accuracy (+30 percentage points over the best fixed-$K$ baseline) via exhaustive retrieval, while IQL matches the best baseline accuracy using 44\% fewer retrieval steps and achieves the only positive episodic return among all policies. Transition-level DPO matches CQL's 92\% accuracy while using 47\% fewer retrieval steps (10.6 vs.\ 20.0), occupying a ``selective-accurate'' region on the Pareto frontier that dominates both CQL and BC. A behavioral cloning baseline matches CQL, confirming that advantage-weighted or preference-based policy extraction is needed to learn selective retrieval. Lambda ablation over step costs $\lambda \in \{0.05, 0.1, 0.2\}$ reveals a clear accuracy-efficiency inflection: only at $\lambda = 0.2$ does CQL transition from exhaustive to selective retrieval.
\end{abstract}

\section{Introduction}

Prior authorization (PA) describes the requirement that certain medical procedures be approved as medically necessary to receive coverage by insurance. These decisions require interpretation of complex policies often using information distributed across multiple documents. Existing retrieval-augmented approaches typically rely on static top-K strategies, retrieving a fixed number of policy sections/chunks \cite{lewis2020}. Such fixed retrieval is fundamentally misaligned with the sequential nature of evidence gathering: it may retrieve unnecessary content, incur avoidable latency and cost, and still miss decisive but lower-ranked sections.

We frame policy retrieval for PA as a sequential decision-making problem \cite{nogueira2017}. We argue for adaptive retrieval, learning both which section to retrieve next and when to stop, to improve efficiency while maintaining accuracy. We model adaptive policy retrieval as a Markov Decision Process (MDP) in which the agent observes the prior authorization request and its retrieval history, selects a candidate policy chunk from a top-K set, or chooses to terminate and issue a determination (approve, deny, or pend). The reward function balances decision correctness against retrieval cost, thus capturing the trade-off between accuracy and efficiency. To ensure safety and reproducibility, we train policies using Conservative Q-Learning (CQL) \cite{kumar2020}, Implicit Q-Learning (IQL) \cite{kostrikov2021}, and Direct Preference Optimization (DPO) \cite{rafailov2023} in an offline reinforcement learning setting \cite{levine2020}, leveraging logged trajectories generated from baseline retrieval strategies over synthetic PA requests derived from publicly available CMS coverage data. This formulation enables the system to learn cost-aware, sequential retrieval strategies while avoiding unsafe exploration in a healthcare context.

\section{Related Work}

Retrieval-augmented generation (RAG) couples parametric models with non-parametric document retrieval, typically retrieving a fixed top-$K$ set of passages before downstream reasoning \cite{lewis2020}. While dense retrieval and passage ranking methods have substantially improved performance on knowledge-intensive tasks, retrieval depth is generally treated as a static hyperparameter rather than an adaptive decision. Recent work has begun to address this limitation: Self-RAG \cite{asai2024} trains a language model to generate reflection tokens that decide whether to retrieve on-demand; FLARE \cite{jiang2023} monitors token-level confidence during generation and triggers retrieval when confidence drops; and Adaptive-RAG \cite{jeong2024} trains a classifier to route queries to different retrieval tiers based on complexity. However, these approaches rely on supervised signals or heuristic thresholds rather than learned cost-aware stopping policies, and none target safety-critical domains where online exploration is impermissible. Previous work has explored RL for information retrieval and document selection, modeling retrieval as a sequential process that can refine queries or select evidence over multiple steps \cite{nogueira2017}. Our work builds on this perspective by explicitly incorporating a stopping decision and optimizing for cost-aware evidence gathering in the healthcare domain using offline RL.

Our training framework draws from offline RL \cite{levine2020} where policies are learned from fixed datasets without environment interaction. Conservative Q-Learning (CQL) penalizes Q-values for out-of-distribution actions, addressing distributional shift \cite{kumar2020}. Direct Preference Optimization (DPO) offers an alternative that directly optimizes the policy using preference comparisons \cite{rafailov2023}, originally for language model alignment but adaptable to sequential MDPs via per-transition preference pairs.

Within healthcare, prior authorization has primarily been addressed through workflow standardization and interoperability initiatives. The HL7 Da Vinci Burden Reduction implementation guides \cite{davinci2024} define FHIR-based APIs for automating PA submission and adjudication, but do not address how a system should consult underlying policy text to gather evidence for a coverage determination. Our work is complementary: we frame policy consultation itself as a learned, cost-aware sequential decision process.

\section{Approach}

Figure~\ref{fig:architecture} shows the system architecture. At each step $t$, the agent observes state $s_t$ (request embedding $\oplus$ retrieval history), selects a chunk from the top-$K$ candidates or stops, and receives a step cost $-\lambda$ or terminal reward $\pm 1$.

\begin{figure}[ht]
\centering
% System Architecture: Adaptive Policy Retrieval MDP
% Usage: \input{figures/system_architecture.tex} inside a figure environment
% Requires in main document preamble:
%   \usepackage{tikz}
%   \usetikzlibrary{arrows.meta,positioning,calc,fit,backgrounds,decorations.pathreplacing}

\definecolor{datablue}{RGB}{66,133,244}
\definecolor{databluefill}{RGB}{219,233,254}
\definecolor{rlgreen}{RGB}{52,168,83}
\definecolor{rlgreenfill}{RGB}{214,243,220}
\definecolor{stopred}{RGB}{214,48,49}
\definecolor{stopredfill}{RGB}{253,228,228}
\definecolor{corpusgold}{RGB}{180,130,30}
\definecolor{corpusgoldfill}{RGB}{255,243,205}
\definecolor{algopurple}{RGB}{120,80,160}

\begin{tikzpicture}[
    >=Stealth,
    node distance=0.4cm and 0.25cm,
    font=\sffamily\scriptsize,
    % Main box style
    box/.style={
        rectangle, rounded corners=2.5pt,
        draw=#1, fill=#1fill, line width=0.5pt,
        minimum height=0.5cm, text width=2.1cm,
        align=center, inner sep=2pt,
    },
    box/.default=datablue,
    % Smaller box style
    sbox/.style={
        rectangle, rounded corners=2pt,
        draw=#1, fill=#1fill, line width=0.4pt,
        minimum height=0.42cm, text width=1.25cm,
        align=center, inner sep=1.5pt,
    },
    sbox/.default=datablue,
    % Wide box style
    wbox/.style={
        rectangle, rounded corners=2.5pt,
        draw=#1, fill=#1fill, line width=0.5pt,
        minimum height=0.5cm, text width=3.0cm,
        align=center, inner sep=2pt,
    },
    wbox/.default=rlgreen,
    % Arrow styles
    arr/.style={->, line width=0.45pt, color=#1!80!black},
    arr/.default=black,
    darr/.style={->, line width=0.4pt, dashed, color=#1!70!black},
    darr/.default=black,
    % Label style
    lbl/.style={font=\sffamily\tiny, color=black!55},
]

% =========================================================
% ROW 1: PA Request
% =========================================================
\node[box] (request) {PA Request\\[-1pt]{\tiny CPT, ICD-10, age}};

% =========================================================
% ROW 2: Encoder
% =========================================================
\node[box, below=of request] (encoder) {S-BERT Encoder\\[-1pt]{\tiny all-MiniLM-L6-v2}};

% =========================================================
% ROW 3: Embeddings (side by side)
% =========================================================
\node[sbox, below left=0.5cm and -0.05cm of encoder] (reqemb)
    {$\mathbf{e}_{\text{req}}$\\[-2pt]{\tiny 384-d}};
\node[sbox, below right=0.5cm and -0.05cm of encoder] (histemb)
    {$\bar{\mathbf{e}}_{\text{hist}}$\\[-2pt]{\tiny 384-d}};

% =========================================================
% ROW 4: State vector
% =========================================================
\node[wbox, below=1.8cm of encoder] (state)
    {State $s_t = [\mathbf{e}_{\text{req}} ;\; \bar{\mathbf{e}}_{\text{hist}}]$\\[-1pt]{\tiny 768-dim vector}};

% =========================================================
% ROW 5: Policy network
% =========================================================
\node[wbox, below=of state] (policy)
    {Policy $\pi_\theta$\\[-1pt]{\tiny MLP: 768$\to$256$\to$256$\to$11}};

% =========================================================
% Training algorithms annotation
% =========================================================
\node[right=0.4cm of policy, font=\sffamily\tiny,
      text width=0.55cm, align=left, color=algopurple] (algos)
    {CQL\\[-1pt]IQL\\[-1pt]DPO\\[-1pt]BC};
\draw[->, solid, color=algopurple, line width=0.4pt] (algos.west) -- (policy.east);

% =========================================================
% ROW 6: Action
% =========================================================
\node[wbox, below=of policy] (action)
    {Action $a_t$};

% =========================================================
% ROW 7: Corpus centered, Retrieve left, Stop right
% =========================================================
% Placed Corpus directly below Action to center it
\node[sbox=corpusgold, below=0.5cm of action] (corpus)
    {Corpus\\[-2pt]{\tiny 186 chunks}};
% Retrieve on the left of Corpus, increased gap to fix top-K overlap
\node[sbox=rlgreen, left=0.8cm of corpus] (retrieve)
    {Retrieve\\[-2pt]{\tiny $a_t \!\in\! \{0\text{-}9\}$}};
% STOP on the right of Corpus, symmetrical gap
\node[sbox=stopred, right=0.8cm of corpus] (stop)
    {STOP\\[-2pt]{\tiny $a_t \!=\! 10$}};

% =========================================================
% ROW 8: Consequences
% =========================================================
\node[box=rlgreen, below=0.4cm of retrieve, text width=1.4cm] (update)
    {Update $\bar{\mathbf{e}}_{\text{hist}}$\\[-1pt]{\tiny $r_t = -\lambda$}};
\node[box=stopred, below=0.4cm of stop, text width=1.4cm] (oracle)
    {Oracle\\[-1pt]{\tiny $R \!\in\! \{+1, -1\}$}};

% =========================================================
% ARROWS: main data flow
% =========================================================
\draw[arr] (request) -- (encoder);
\draw[arr] (encoder.south) -- ++(0,-0.12) -| (reqemb.north);

\draw[arr=datablue] (reqemb.south) -- ++(0,-0.5)
    -| ($(state.north west)!0.3!(state.north east)$);
\draw[arr=datablue] (histemb.south) -- ++(0,-0.5)
    -| ($(state.north west)!0.7!(state.north east)$);

% =========================================================
% ARROWS: RL loop
% =========================================================
\draw[arr=rlgreen] (state) -- (policy);
\draw[arr=rlgreen] (policy) -- (action);

% Branching arrows from Action to Retrieve and STOP (routing cleanly over Corpus)
\draw[arr=rlgreen] (action.south) -- ++(0,-0.15) -| (retrieve.north)
    node[pos=0.25, left, lbl] {};
\draw[arr=stopred] (action.south) -- ++(0,-0.15) -| (stop.north)
    node[pos=0.25, right, lbl] {};

\draw[arr=rlgreen] (retrieve) -- (update);
\draw[arr=stopred] (stop) -- (oracle);

% =========================================================
% Corpus to retrieve
% =========================================================
\draw[arr=corpusgold] (corpus.west) -- (retrieve.east)
    node[midway, above, lbl] {top-$K$};

% =========================================================
% LOOP-BACK: update -> history embedding (MDP transition)
% =========================================================
% Routes from Update (left), strictly underneath the whole structure, to the right side
\coordinate (farRight) at ($(oracle.east) + (0.35, 0)$);
\draw[arr=rlgreen, rounded corners=3pt]
    (update.south) -- ++(0,-0.4) coordinate (U)
    -- (U -| farRight) coordinate (V)
    -- (V |- histemb.east) node[midway, left, lbl] {$s_{t+1}$}
    -- (histemb.east);

% =========================================================
% Brace: concat hint
% =========================================================
\draw[decorate, decoration={brace, amplitude=2pt, mirror},
      line width=0.3pt, black!35]
    ($(reqemb.south west)+(0.02,-0.18)$) -- ($(histemb.south east)+(-0.02,-0.18)$)
    node[midway, below=2.5pt, lbl] {concatenate};

\end{tikzpicture}
\caption{System architecture. The agent iteratively retrieves policy chunks from a Sentence-BERT-indexed corpus and decides when to stop. The oracle evaluates the PA request using only retrieved chunks.}
\label{fig:architecture}
\end{figure}

\subsection{Data \& Simulator}

\begin{itemize}
\item \textbf{Policy Corpus \& Retrieval.} We constructed a corpus of 186 policy chunks from the CMS Medicare Coverage Database (MCD), spanning 10 medical procedures (Table~\ref{tab:corpus}). Source documents include Local Coverage Determinations (LCD), National Coverage Determinations (NCD), and Local Coverage Articles (LCA), parsed into paragraph-level chunks. Each chunk is encoded into a 384-dimensional vector using Sentence-BERT (\textit{all-MiniLM-L6-v2}). At the corpus level, 31\% of chunks are shared across procedures, and billing documentation from one procedure can rank higher than coverage-specific chunks for another due to shared administrative terminology. This cross-procedure semantic interference creates a realistic retrieval challenge. The simulator returns the top-$K$ chunks by cosine similarity not yet selected.
\item \textbf{Synthetic Request Generation.} A rule-based generator creates synthetic PA requests for 10 CMS procedures (Table~\ref{tab:corpus}). For each procedure, it samples a diagnosis code from the procedure's ICD-10 set, a patient age from a clinically plausible range, and a ground-truth outcome (approve, deny, or pend) with balanced distribution. These fields are encoded via Sentence-BERT to produce the query vector $e_q$. Procedures span a range of retrieval difficulty: easy ones (Colonoscopy, Mammography) have distinctive terminology, while hard ones (CT Head, MRI Lumbar) suffer from cross-procedure interference.
\item \textbf{Oracle \& Evaluation.} A rule-based oracle establishes ground truth by evaluating each request against the full corpus. When the agent stops, the oracle re-evaluates using only retrieved chunks. The agent succeeds if it retrieved sufficient evidence for the oracle to reach the correct decision.
\end{itemize}

\textbf{Concrete episode walkthrough.} Consider a PA request for \textit{MRI Lumbar Spine} (CPT 72148) for a 55-year-old patient with diagnosis L54 (dorsalgia). The request generator produces structured metadata (procedure code, diagnosis, age, prior treatments), which the retriever encodes via Sentence-BERT and scores against all 186 corpus chunks by cosine similarity. The agent observes the initial state $s_0$ (request embedding $\oplus$ zero vector) and iteratively selects chunks. Suppose the agent retrieves 3 chunks: a coverage-criteria chunk referencing indications for lumbar MRI, a documentation-requirements chunk listing clinical prerequisites, and a billing-article chunk on CPT 72148 coding. After each retrieval step, the state updates via mean-pooling, incurring cost $-\lambda$. The agent then selects the stop action ($a = K$). The oracle evaluates the request using \textit{only these 3 retrieved chunks}: it checks whether the retrieved evidence contains a matching diagnosis code and coverage criteria. If the oracle's decision (e.g.\ ``approve'') matches the ground-truth oracle decision (computed using the full corpus), the agent receives $r_T = +1$; otherwise $r_T = -1$. The total return for this episode is $G = +1 - 3\lambda$.

\textbf{Corpus composition.} Table~\ref{tab:corpus} summarizes per-procedure statistics. By type: 107 coverage-criteria, 71 billing, and 8 exclusion chunks. Shared-chunk ratios range from 0\% (Colonoscopy, Speech-Language) to 77--100\% (imaging procedures), driving the cross-procedure interference that makes imaging procedures hardest (Figure~\ref{fig:per_proc}).

\textbf{Dataset splits.} We generated 2{,}000 training and 200 test episodes, stratified across procedures. Ground-truth decisions: 37.6\% approve, 11.9\% deny, 50.5\% pend.

\begin{table}[h]
\centering
\caption{Corpus and dataset composition per procedure. \textit{Shared} = chunks appearing in multiple procedures' retrieval pools.}
\label{tab:corpus}
% Reduce space between columns (default is 6pt)
\setlength{\tabcolsep}{2pt} 
\begin{small}
\begin{sc}
\begin{tabular}{lcccc}
\toprule
Procedure & Chunks & Shared & Train & Test \\
\midrule
45378 (Colonoscopy) & 30 & 0 & 201 & 23 \\
70450 (CT Head) & 56 & 54 & 195 & 27 \\
70486 (CT Maxfac.) & 43 & 43 & 199 & 19 \\
70553 (MRI Brain) & 54 & 54 & 208 & 28 \\
71260 (CT Chest) & 44 & 42 & 205 & 20 \\
72148 (MRI Lumbar) & 55 & 49 & 178 & 20 \\
74177 (CT Abdomen) & 46 & 42 & 193 & 12 \\
77067 (Mammography) & 41 & 31 & 191 & 18 \\
92507 (Speech-Lang.) & 48 & 0 & 217 & 13 \\
92550 (Tympanometry) & 58 & 31 & 213 & 20 \\
\midrule
\textbf{Total} & \textbf{186} & \textbf{57} & \textbf{2000} & \textbf{200} \\
\bottomrule
\end{tabular}
\end{sc}
\end{small}
\end{table}

\subsection{MDP Formulation}

We formulate adaptive policy retrieval as a finite-horizon Markov Decision Process $(S, A, P, R, \gamma, \rho_0)$, where $S$ is the state space, $A$ the action space, $P(s' | s, a)$ the transition kernel, $R(s,a)$ the reward function, $\gamma$ the discount factor, and $\rho_0$ the initial state distribution over PA requests. Episodes run for at most $H = 20$ steps.

\begin{itemize}
\item \textbf{States.} The state $s_t \in \mathbb{R}^{768}$ concatenates two 384-dimensional components: (1) a request embedding from the PA request text via a sentence-transformer encoder, and (2) the element-wise mean of all chunk embeddings retrieved so far (zero if none). This captures both \textit{what the agent is looking for} and \textit{what it knows so far}, maintaining constant dimensionality and Markovianity.
\item \textbf{Actions.} The agent selects from $A = \{0, 1, \dots, K\}$: actions $0$ through $K\!-\!1$ select the corresponding candidate from a top-$K$ list ranked by cosine similarity (action semantics are state-dependent), and action $K$ stops retrieval. We use $K = 10$.
\item \textbf{Transitions.} Deterministic: selecting $a \in \{0, \dots, K\!-\!1\}$ appends the chunk, updates the state via mean-pooling, and refreshes the candidate list. The stop action or reaching $H$ terminates the episode.
\item \textbf{Rewards.} Each retrieval incurs cost $r_t = -\lambda$ ($\lambda > 0$, ablated over $\{0.05, 0.1, 0.2\}$). Upon stopping, the oracle evaluates using only retrieved chunks; the terminal reward is $r_T = +1$ if the decision matches ground truth, $-1$ otherwise. Total return: $G = r_T - \lambda \cdot n$. We use $\gamma = 1.0$, directly trading off correctness against cumulative retrieval cost.
\end{itemize}

\subsection{Conservative Q-Learning}

All algorithms are implemented from scratch in PyTorch without external RL libraries. We train the retrieval policy using Conservative Q-Learning \cite{kumar2020}, an offline reinforcement learning algorithm. Offline RL is necessary because we cannot deploy untrained agents in a healthcare setting to collect online data; instead, we train on a fixed dataset of episodes collected by behavior policies (uniform random, fixed-K retrieval, and a heuristic policy).

\textbf{Q-Network.} We approximate the action-value function $Q(s, a)$ with a 3-layer multi-layer perceptron: 768 (input) to 256 (ReLU) to 256 (ReLU) to 11 (output, no activation). The network has 265,483 trainable parameters. We maintain two copies of the network: a main network updated at every training step, and a target network synchronized every 10 epochs via hard update for stable Bellman targets.

\textbf{Loss Function.} The Conservative Q-Learning loss consists of two components:
\begin{equation}
L = L_{TD} + \alpha \cdot L_{cons}
\end{equation}
The temporal difference loss minimizes the squared Bellman error:
\begin{equation}
\begin{aligned}
L_{TD} &= \mathbb{E}_{(s,a,r,s')\sim D} [ (Q(s,a) \\
&\quad - (r + \gamma \max_{a'} Q_{target}(s', a')))^2 ]
\end{aligned}
\end{equation}
The conservative penalty discourages overestimation of out-of-distribution actions:
\begin{equation}
\begin{aligned}
L_{cons} &= \mathbb{E}_{s\sim D} \Big[ \log \sum_a \exp(Q(s,a)) \\
&\quad - \mathbb{E}_{a\sim D} [Q(s, a)] \Big]
\end{aligned}
\end{equation}
The coefficient $\alpha$ controls the degree of conservatism. We use $\alpha = 1.0$ as the initial default and sweep over $\{0.1, 0.5, 1.0\}$. This penalty is critical in our setting because the offline dataset covers only a small fraction of the continuous state space, making standard Q-learning prone to overestimating Q-values for state-action pairs not seen during data collection.

\textbf{Training.} We train using Adam (learning rate $3\times10^{-4}$) with gradient clipping (max norm 1.0) for 200 epochs. Each epoch samples a batch of 256 transitions uniformly from the replay buffer. We use an undiscounted formulation ($\gamma = 1.0$) because episodes are short (at most 20 steps). Training metrics (loss, TD error, conservative penalty, mean Q-values) are logged for convergence monitoring. The total loss converges from 2.62 to 1.71 over training.

\subsection{Implicit Q-Learning}

As a second offline RL algorithm, we train the retrieval policy using Implicit Q-Learning (IQL) \cite{kostrikov2021}. While CQL addresses distributional shift by explicitly penalizing Q-values for out-of-distribution actions, IQL takes a different approach: it avoids querying the value of unseen actions altogether. Instead of computing $\max_{a'} Q(s', a')$ in the Bellman target, IQL learns a separate state-value function $V(s)$ that approximates the value of good actions through expectile regression. This makes training more stable because the Q-network never needs to evaluate actions outside the dataset. Including IQL alongside CQL lets us compare two philosophically different strategies for offline RL on the same retrieval task.

\textbf{Architecture.} IQL maintains four networks (768$\to$256$\to$256): Q-network $Q_\theta$ (11 outputs), target $Q_{\bar{\theta}}$ (hard update every 10 epochs), value network $V_\psi$ (scalar), and policy $\pi_\phi$ (action logits).

\textbf{Loss Functions.} IQL training consists of three coupled losses. The value network is trained via asymmetric expectile regression:
\begin{equation}
    L_V = \mathbb{E}_{(s,a) \sim D} \left[ L_2^\tau \left( Q_{\bar{\theta}}(s, a) - V_\psi(s) \right) \right]
\end{equation}
where $L_2^\tau(u) = |\tau - \mathbf{1}(u < 0)| \cdot u^2$ is the asymmetric squared loss. Setting $\tau > 0.5$ biases $V$ toward the upper expectile of $Q$, so $V$ learns to approximate the value of better-than-average actions without ever maximizing over unseen actions. We use $\tau = 0.9$.

The Q-network is updated via standard TD learning, but uses $V(s')$ instead of $\max_{a'} Q(s', a')$ as the bootstrap target:
\begin{equation}
    L_Q = \mathbb{E}_{(s,a,r,s') \sim D} \left[ \left( Q_\theta(s, a) - (r + \gamma V_\psi(s')) \right)^2 \right]
\end{equation}
This avoids the maximization step that causes overestimation in standard offline Q-learning.

Finally, the policy is extracted through advantage-weighted behavioral cloning:
\begin{equation}
    L_\pi = -\mathbb{E}_{(s,a) \sim D} \left[ \exp(\beta \cdot A(s, a)) \cdot \log \pi_\phi(a \mid s) \right]
\end{equation}
where $A(s, a) = Q_{\bar{\theta}}(s, a) - V_\psi(s)$ is the advantage and $\beta$ is an inverse temperature that controls how strongly the policy concentrates on high-advantage actions. Actions with positive advantage receive exponentially higher weight in the supervised loss, while low-advantage actions are effectively down-weighted. The advantage weights are clamped at 100.0 to prevent numerical overflow.

\textbf{Training.} All three networks are trained jointly using Adam (learning rate $10^{-3}$) with gradient clipping (max norm 1.0) for 1000 epochs. Each epoch samples a batch of 256 transitions from the shared replay buffer. We use $\tau = 0.9$ and $\beta = 10.0$, empirically tuned for the PA retrieval environment; the Kostrikov et al.\ defaults ($\tau = 0.7$, $\beta = 3.0$, 200 epochs) underperform on this corpus. At inference time, the learned policy network $\pi_\phi$ selects the action with the highest logit, with invalid actions masked out.

\subsection{Behavioral Cloning Baseline}

We include a behavioral cloning (BC) baseline \cite{pomerleau1991} to test whether RL adds value beyond imitation. BC minimizes cross-entropy $L_{BC} = -\mathbb{E}_{(s, a) \sim D} [ \log \pi_\phi(a \mid s) ]$ using no reward information, no bootstrapping, and no temporal difference learning. It reuses the same MLP architecture (768$\to$256$\to$256$\to$11) and is trained with Adam (lr $10^{-3}$) for 100 epochs.

\subsection{Direct Preference Optimization}
\label{sec:dpo}

As a fourth training algorithm, we adapt Direct Preference Optimization (DPO) \cite{rafailov2023} from language model alignment to our sequential MDP. DPO sidesteps reward modeling by directly optimizing the policy on preference pairs: given two behaviors, the policy learns to assign higher probability to the preferred one. In our setting, preference pairs are constructed from the offline dataset by comparing episodes with different returns.

\textbf{Transition-level formulation.} Standard DPO operates at the trajectory level, comparing full-episode log-probabilities. We found that trajectory-level DPO cannot extrapolate beyond the episode lengths present in the training data (3--8 steps), which is critical because evaluation runs for up to 20 steps. We therefore adopt a transition-level formulation analogous to token-level DPO in language models: preference pairs are constructed per retrieval step rather than per episode. Given a winner episode (higher return) and a loser episode (lower return), we pair their actions at each shared retrieval depth $t$, producing state-action preference tuples $(s_t^w, a_t^w, s_t^l, a_t^l)$. This preserves gradient signal at each step and enables generalization to longer episodes at evaluation time.

\textbf{Loss function.} The DPO objective minimizes:
\begin{equation}
    L_{DPO} = -\mathbb{E}_{(s^w, a^w, s^l, a^l)} \left[ \log \sigma \left( \beta \cdot \Delta \right) \right]
\end{equation}
where $\Delta = \log \frac{\pi_\theta(a^w \mid s^w)}{\pi_{ref}(a^w \mid s^w)} - \log \frac{\pi_\theta(a^l \mid s^l)}{\pi_{ref}(a^l \mid s^l)}$ compares the log-probability ratios between the learned policy $\pi_\theta$ and a frozen reference policy $\pi_{ref}$, and $\beta$ controls the strength of the KL constraint.

\textbf{Reference policy and BC warmup.} The reference policy $\pi_{ref}$ is initialized via behavioral cloning for 200 epochs on the offline buffer, then frozen. A well-converged reference is critical: if $\pi_{ref}$ is near-random, the KL constraint carries no useful information and DPO fine-tuning degrades. We verified BC warmup convergence (loss plateau at 0.88) before freezing.

\textbf{Architecture and training.} DPO reuses the same 3-layer MLP (768$\to$256$\to$256$\to$11) as CQL, IQL, and BC, maintaining two copies: the trainable policy $\pi_\theta$ (initialized from $\pi_{ref}$) and the frozen reference $\pi_{ref}$. DPO is trained with Adam (lr $10^{-4}$) for 2,000 epochs with $\beta = 3.0$, gradient clipping at 1.0, and batch size 256. Hyperparameters were selected via a sweep over $\beta \in \{0.5, 1.0, 3.0\}$ (Section~\ref{sec:dpo_beta}).

\section{Experiments \& Results}

\subsection{Experimental Setup}

We train all agents on 2,000 offline episodes (8,352 transitions) from a mixture of behavior policies: FixedK($k\!\in\!\{3,5\}$), Heuristic(0.8), and two epsilon-greedy variants. Each policy is evaluated on 200 held-out episodes.

\subsection{Main Results}

Table \ref{tab:main_results} presents on-policy evaluation results on the 186-chunk corpus with 10 CMS procedures.

\begin{table}[ht]
\caption{On-policy evaluation (200 episodes, 186 chunks)}
\label{tab:main_results}
\vskip 0.1in
\begin{center}
\begin{small}
\begin{sc}
\begin{tabular}{lccc}
\toprule
Policy & Accuracy & Return & Steps \\
\midrule
\textbf{DPO (transition)} & \textbf{92.0\%} & $\mathbf{-0.12}$ & \textbf{10.6} \\
CQL & 92.0\% & $-1.06$ & 20.0 \\
BC (imitation) & 92.0\% & $-1.06$ & 20.0 \\
IQL & 62.5\% & $+0.01$ & 3.4 \\
FixedK($k=5$) & 62.0\% & $-0.26$ & 6.0 \\
FixedK($k=3$) & 54.0\% & $-0.22$ & 4.0 \\
Heuristic(0.8) & 51.0\% & $-0.22$ & 3.4 \\
\bottomrule
\end{tabular}
\end{sc}
\end{small}
\end{center}
\vskip -0.1in
\end{table}

CQL achieves 92.0\% accuracy, a 30 percentage-point improvement over the best fixed baseline FixedK($k\!=\!5$) at 62.0\%. However, CQL achieves this by exhaustively retrieving all 20 steps in every episode, resulting in the highest retrieval cost and a negative return of $-1.06$.

IQL presents a contrasting strategy: it achieves 62.5\% accuracy (matching FixedK($k\!=\!5$)) in only 3.4 retrieval steps, 44\% fewer steps than the baseline's 6.0. IQL is the only policy with positive episodic return ($+0.01$), meaning it is the only policy where correct decisions outweigh retrieval costs.

Transition-level DPO matches CQL's 92.0\% accuracy while using only 10.6 retrieval steps, a 47\% reduction. Its return of $-0.12$ is 89\% better than CQL's $-1.06$, establishing DPO as the strongest overall policy: it achieves maximum accuracy with near-minimal retrieval cost. The per-step formulation enables DPO to generalize from the training data's 3--8 step episodes to 20-step evaluation episodes, unlike trajectory-level DPO which is bounded by the episode lengths observed during training (see Section~\ref{sec:dpo}).

Behavioral cloning (BC) exactly matches CQL at 92.0\% accuracy and 20.0 steps. Since BC learns to imitate the behavior policy without any reward signal, this shows that CQL's conservative penalty reinforces the exhaustive retrieval pattern already present in the training data rather than discovering a novel strategy. IQL and DPO are the only algorithms that learn qualitatively different (selective) retrieval strategies: IQL through advantage-weighted policy extraction, and DPO through per-transition preference learning.

\subsection{Lambda Ablation}

Table \ref{tab:lambda} shows the effect of varying the step cost $\lambda$ on CQL training.

\begin{table}[ht]
\caption{CQL lambda ablation (200 episodes)}
\label{tab:lambda}
\vskip 0.1in
\begin{center}
\begin{small}
\begin{sc}
\begin{tabular}{lccc}
\toprule
$\lambda$ & Accuracy & Steps & Behavior \\
\midrule
0.05 & 92.0\% & 20.0 & Exhaustive \\
0.1 (default) & 92.0\% & 20.0 & Exhaustive \\
\textbf{0.2} & \textbf{91.5\%} & \textbf{14.9} & Selective \\
\bottomrule
\end{tabular}
\end{sc}
\end{small}
\end{center}
\vskip -0.1in
\end{table}

At $\lambda = 0.05$ and $\lambda = 0.1$, the step cost is insufficient to overcome the accuracy benefit of exhaustive retrieval: the penalty for 20 steps ($-20\lambda = -1.0$ at $\lambda = 0.05$) is outweighed by the $+1.0$ terminal reward from correct decisions. At $\lambda = 0.2$, the penalty per step doubles, making exhaustive retrieval too expensive. CQL learns selective retrieval, reducing steps from 20.0 to 14.9 while losing only 0.5 percentage points of accuracy. Each lambda value requires a separately collected offline dataset because the reward signal (including step cost) is baked into the replay buffer at collection time.

\begin{figure}[ht]
\centering
\includegraphics[width=0.48\textwidth]{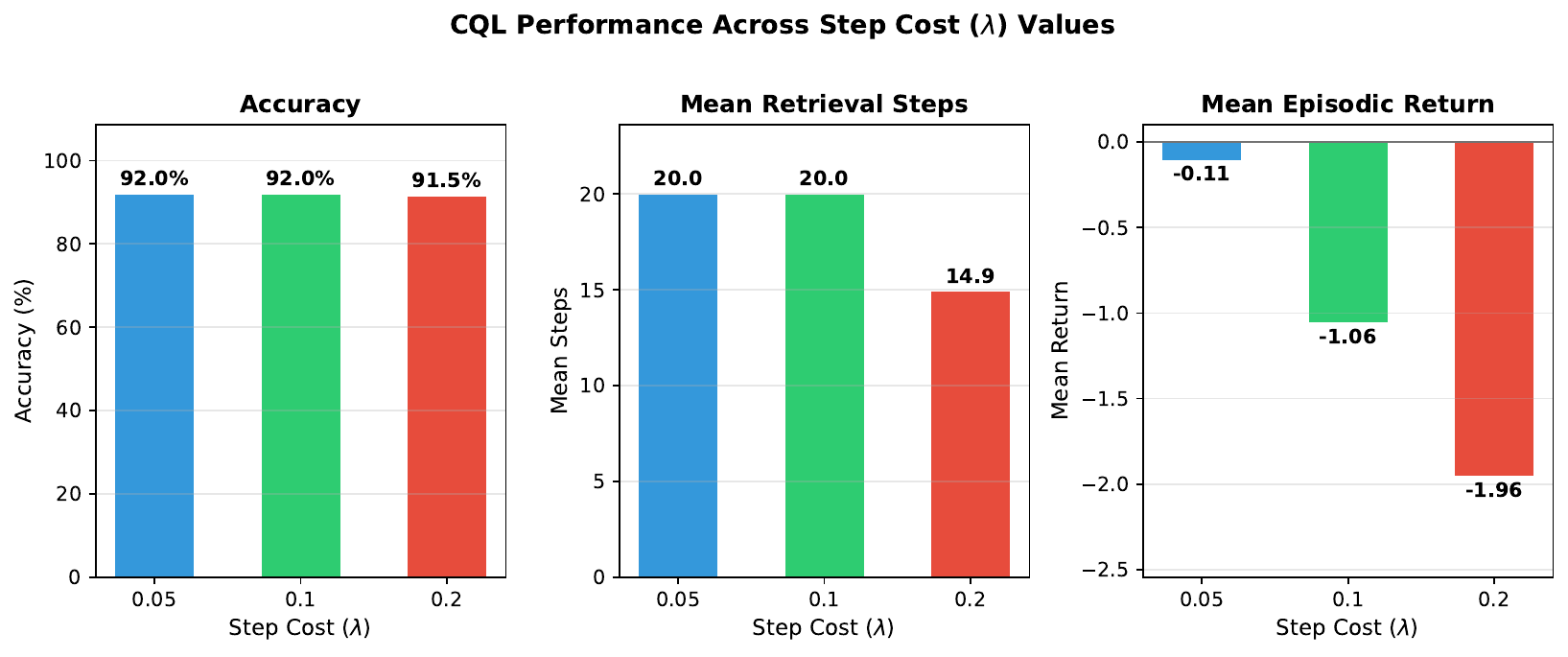}
\caption{Lambda ablation: accuracy, mean steps, and mean return for CQL across step cost values $\lambda \in \{0.05, 0.1, 0.2\}$. At $\lambda = 0.2$, CQL transitions from exhaustive to selective retrieval.}
\label{fig:lambda}
\end{figure}

\subsection{DPO Beta Ablation}
\label{sec:dpo_beta}

Table~\ref{tab:dpo_beta} shows the effect of $\beta$ (the KL constraint strength) on DPO's accuracy-efficiency tradeoff. All configurations use transition-level DPO with 200 BC warmup epochs.

\begin{table}[ht]
\caption{DPO beta ablation (200 episodes, transition-level)}
\label{tab:dpo_beta}
\vskip 0.1in
\begin{center}
\begin{small}
\begin{sc}
\begin{tabular}{lcccc}
\toprule
$\beta$ & Epochs & Accuracy & Return & Steps \\
\midrule
0.5 & 500 & 79.5\% & $-0.45$ & 11.4 \\
1.0 & 1000 & 85.5\% & $-0.31$ & 11.2 \\
\textbf{3.0} & \textbf{2000} & \textbf{92.0\%} & $\mathbf{-0.12}$ & \textbf{10.6} \\
\bottomrule
\end{tabular}
\end{sc}
\end{small}
\end{center}
\vskip -0.1in
\end{table}

Unlike CQL's lambda ablation, which trades accuracy for efficiency, DPO's beta sweep reveals monotonic improvement: higher $\beta$ improves both accuracy and return simultaneously while maintaining stable retrieval depth ($\approx$11 steps). This is because higher $\beta$ strengthens the KL constraint, keeping the policy closer to the well-converged BC reference and preventing distribution collapse. At $\beta = 3.0$, DPO matches CQL's 92\% accuracy while using roughly half the retrieval steps.

\subsection{Convergence Analysis}

Both CQL and IQL training converge successfully with no NaN or divergence detected. CQL total loss decreases from 2.62 to 1.71 (35\% reduction over 200 epochs). The conservative penalty decreases from 2.38 to 1.21 (49\% reduction), while TD loss increases from 0.24 to 0.49; this is expected behavior, as the conservative penalty shifts Bellman targets. IQL Q-network TD loss converges from 0.24 to 0.01 (95\% reduction over 1000 epochs), and policy loss from 2.06 to 0.94 (55\% reduction). DPO converges in two phases: BC warmup loss from 2.38 to 0.88 (63\% reduction over 200 epochs), followed by DPO loss from 0.69 to 0.35 (49\% reduction over 2000 epochs) with preference accuracy increasing from 28\% to 85\%.

\begin{figure}[ht]
\centering
\includegraphics[width=0.48\textwidth]
{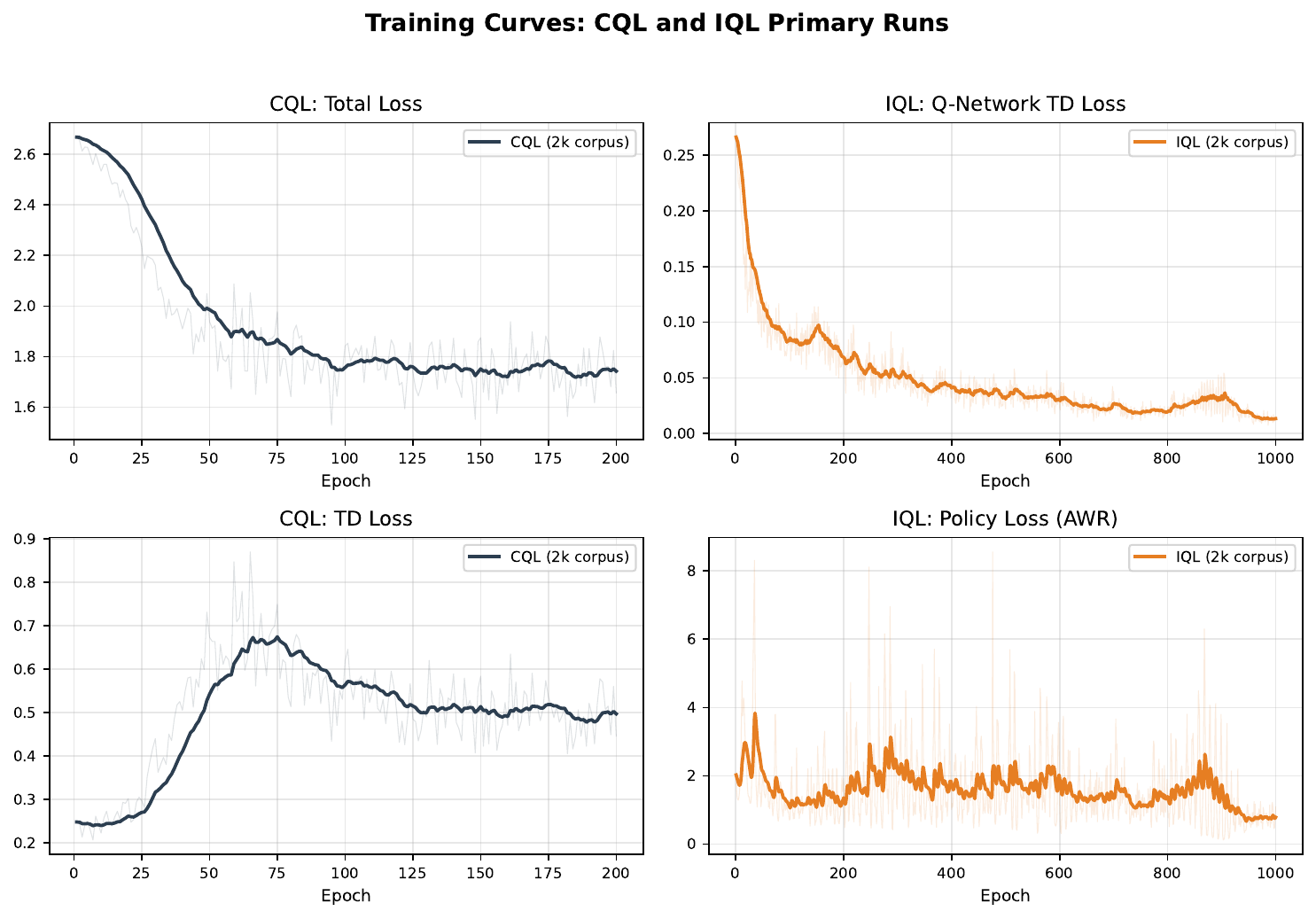}
\caption{Training curves for CQL (200 epochs) and IQL (1000 epochs). Left: CQL total loss and TD loss. Right: IQL Q-network TD loss and policy loss. All metrics converge without divergence.}
\label{fig:training_curves}
\end{figure}

\begin{figure}[ht]
\centering
\includegraphics[width=0.48\textwidth]{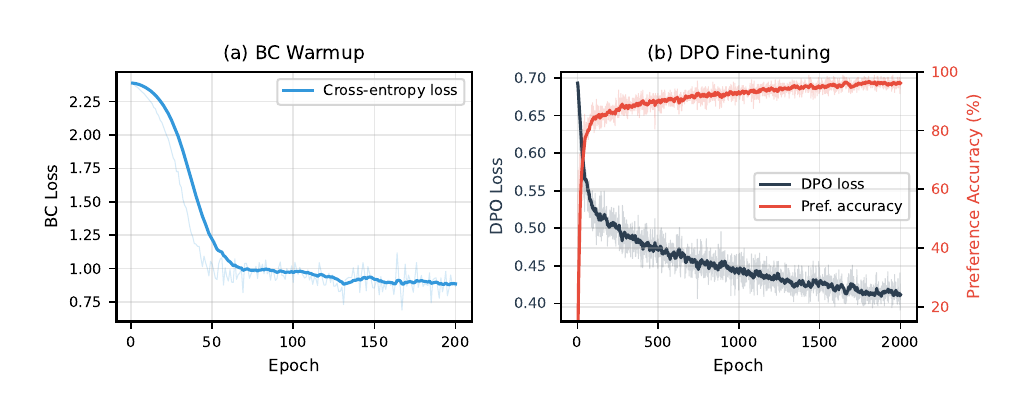}
\caption{DPO training curves. Left: BC warmup cross-entropy loss (200 epochs), converging from 2.38 to 0.88. Right: DPO loss (blue) and preference accuracy (red) over 2000 epochs. Preference accuracy reaches 85\%, indicating the policy reliably assigns higher probability to preferred actions.}
\label{fig:dpo_training}
\end{figure}

\subsection{Off-Policy Evaluation}

We evaluate all four trained policies using Weighted Importance Sampling (WIS, ratios clipped to $[0.01, 100]$) and Fitted Q-Evaluation (FQE, 200 epochs per policy) on the training buffer.

\begin{table}[ht]
\caption{Off-policy evaluation estimates}
\label{tab:ope}
\vskip 0.1in
\begin{center}
\begin{small}
\begin{sc}
\begin{tabular}{lcc}
\toprule
Policy & WIS & FQE Mean $\hat{Q}$ \\
\midrule
CQL & $-0.055$ & $-2.22$ \\
DPO & $\mathbf{-0.012}$ & $\mathbf{-0.34}$ \\
IQL & $-0.021$ & $-0.17$ \\
BC & $-0.084$ & $-2.41$ \\
\bottomrule
\end{tabular}
\end{sc}
\end{small}
\end{center}
\vskip -0.1in
\end{table}

IQL achieves the best FQE estimate (least negative $\hat{Q}$), consistent with its positive on-policy return. DPO achieves the best WIS estimate ($-0.012$) and second-best FQE ($-0.34$), reflecting its combination of high accuracy and moderate retrieval cost. CQL and BC show similar off-policy estimates ($\approx -2.3$ FQE), reinforcing the finding that CQL's policy is behaviorally equivalent to imitation. The rank ordering across OPE methods agrees with on-policy evaluation.

\subsection{Statistical Significance}

We perform paired t-tests on per-episode correctness indicators between policies (Table~\ref{tab:significance}). DPO and CQL both significantly outperform all baselines at $p < 0.001$, but DPO vs.\ CQL shows no significant accuracy difference ($p = 1.0$), with DPO achieving the same accuracy at substantially lower retrieval cost.

\begin{table}[ht]
\caption{Statistical significance: pairwise comparisons}
\label{tab:significance}
% Reduce space between columns (default is 6pt)
\setlength{\tabcolsep}{4pt} 
\vskip 0.1in
\begin{center}
\begin{small}
\begin{sc}
\begin{tabular}{lccc}
\toprule
Comparison & $\Delta$ Acc & $p$-value & 95\% CI \\
\midrule
DPO vs CQL & 0.0pp & -- & [0.0, 0.0] \\
DPO vs IQL & +29.5pp & $8.1 \times 10^{-17}$ & [23.5, 36.0] \\
DPO vs BC & 0.0pp & -- & [0.0, 0.0] \\
CQL vs IQL & +29.5pp & $8.1 \times 10^{-17}$ & [23.5, 36.0] \\
CQL vs BC & 0.0pp & -- & [0.0, 0.0] \\
CQL vs FixedK(3) & +38.0pp & $2.0 \times 10^{-22}$ & [31.5, 45.0] \\
CQL vs FixedK(5) & +30.0pp & $3.9 \times 10^{-17}$ & [23.5, 36.5] \\
CQL vs Heuristic & +41.0pp & $1.4 \times 10^{-24}$ & [34.0, 48.0] \\
\bottomrule
\end{tabular}
\end{sc}
\end{small}
\end{center}
\vskip -0.1in
\end{table}

\subsection{Accuracy-Efficiency Tradeoff}

\begin{figure}[ht]
\centering
\includegraphics[width=0.48\textwidth]{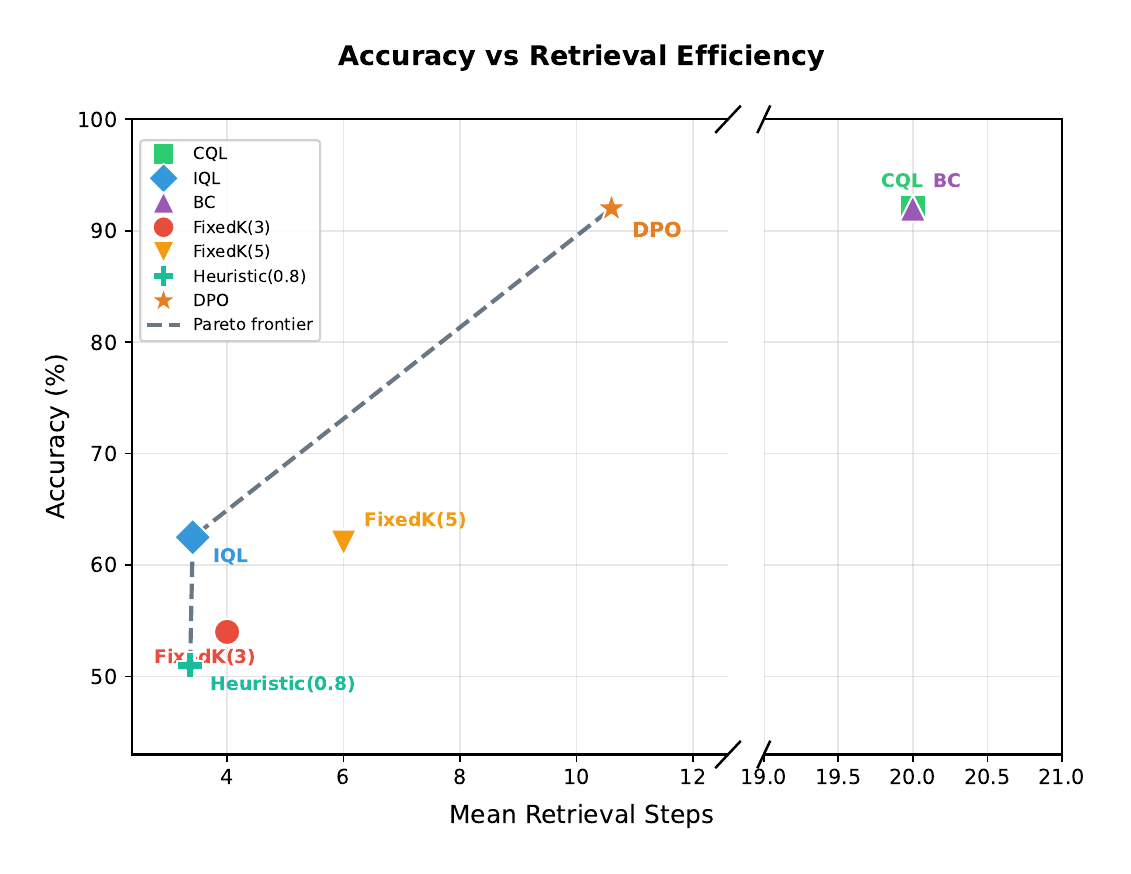}
\caption{Accuracy vs.\ retrieval steps for all seven policies. DPO achieves (10.6, 92\%) on the Pareto frontier, dominating CQL and BC at (20, 92\%). IQL achieves (3.4, 62.5\%).}
\label{fig:pareto}
\end{figure}

Figure~\ref{fig:pareto} presents the Pareto frontier over all seven policies. CQL and BC occupy the top-right corner (high accuracy, high cost), while IQL sits on the Pareto frontier alongside FixedK($k\!=\!5$) but achieves the same accuracy in 44\% fewer steps. DPO occupies a ``selective-accurate'' region of the frontier at (10.6, 92\%), achieving the same accuracy as CQL/BC with 47\% fewer retrieval steps. This position dominates both CQL and BC (same accuracy, lower cost), making them Pareto-suboptimal. The three-point Pareto frontier (IQL, DPO, and the origin) demonstrates that preference-based and advantage-weighted methods can learn retrieval strategies that value-based methods cannot.

\subsection{Per-Procedure Analysis}
\label{sec:per_proc}

\begin{figure}[ht]
\centering
\includegraphics[width=0.48\textwidth]{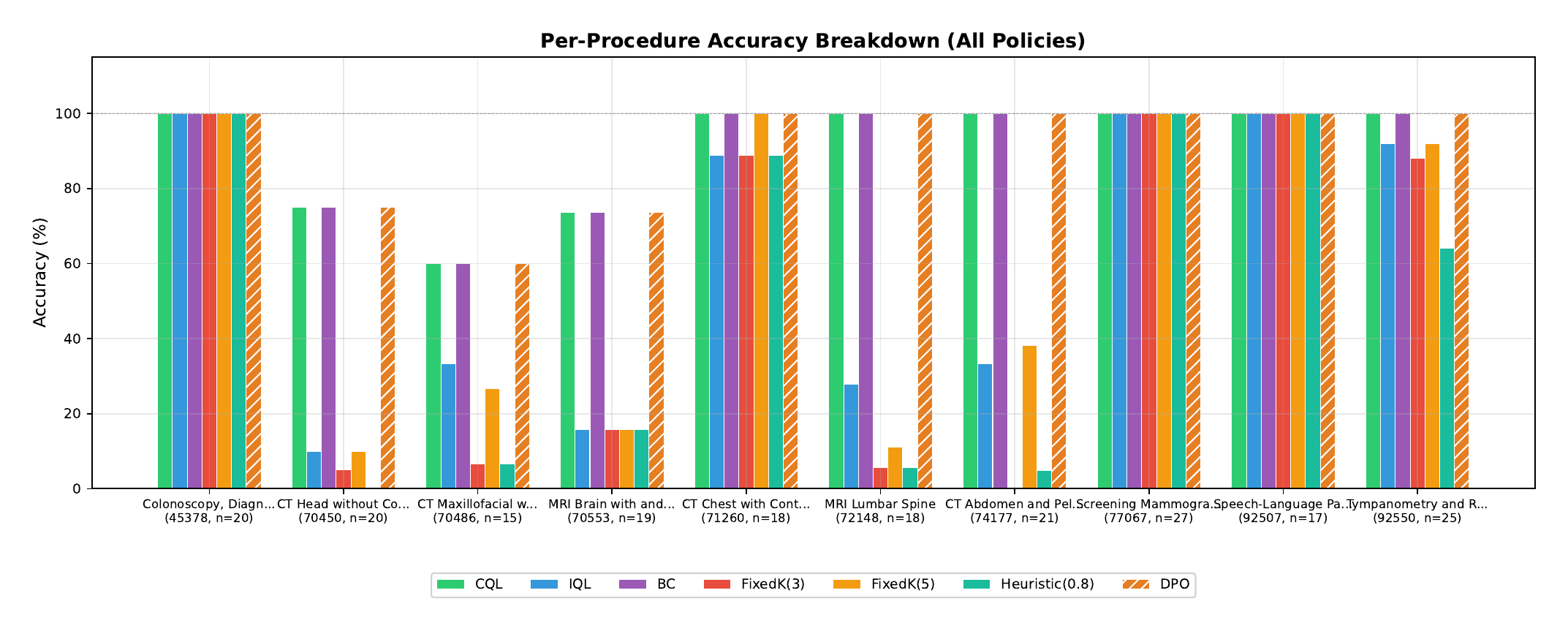}
\caption{Per-procedure accuracy breakdown for all seven policies. CQL, BC, and DPO achieve 100\% on 7/10 procedures; three imaging procedures with high cross-procedure semantic interference remain challenging for all policies.}
\label{fig:per_proc}
\end{figure}

Figure~\ref{fig:per_proc} reveals that aggregate accuracy masks substantial per-procedure variation. CQL, BC, and DPO all achieve 100\% accuracy on 7 of 10 procedures, with the three hardest procedures -- CT Head (70450), CT Maxillofacial (70486), and MRI Brain (70553) -- driving their overall rates. These hard procedures share high cross-procedure semantic interference, where billing and administrative chunks from other procedures crowd out coverage-specific evidence. DPO matches CQL and BC on every procedure (75\%, 60\%, and 74\% on the hard three) while using fewer retrieval steps, confirming that its selective strategy sacrifices no per-procedure accuracy.

IQL accuracy drops most sharply on these same hard procedures (10\%, 33\%, 16\% respectively), confirming that aggressive early stopping is most vulnerable to corpus complexity. On easy procedures (Colonoscopy, Mammography, Speech-Language), all policies achieve 100\%.

\begin{figure}[ht]
\centering
\includegraphics[width=0.48\textwidth]{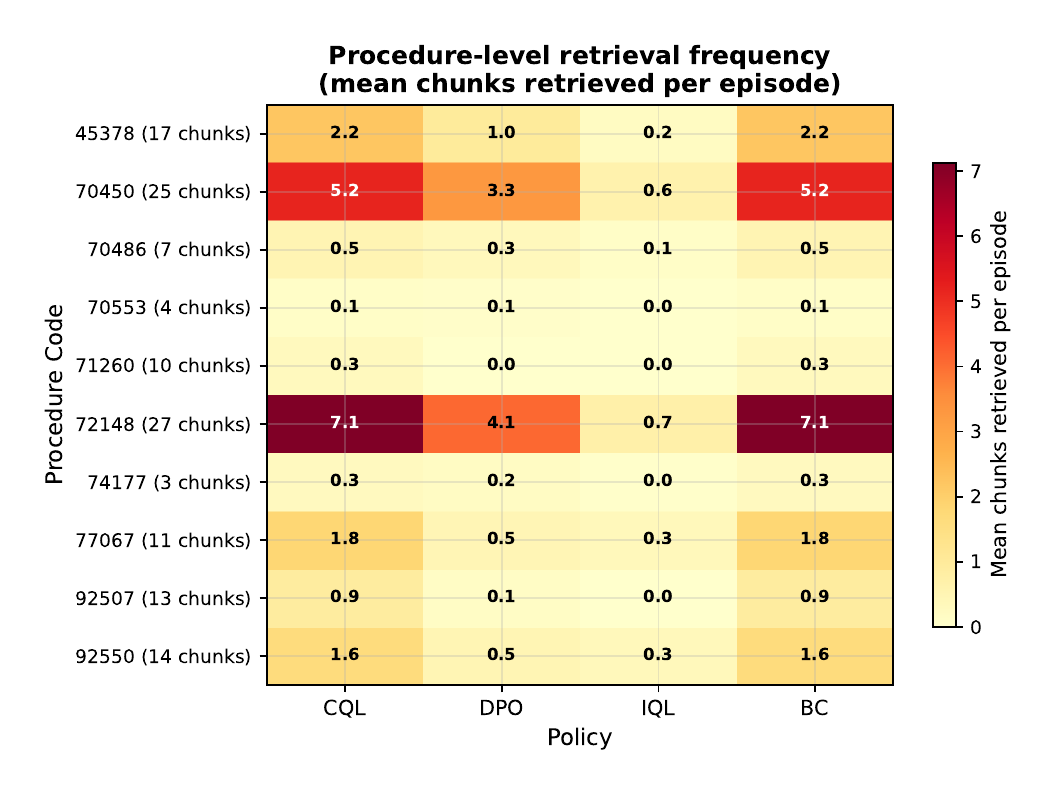}
\caption{Procedure-level retrieval frequency (mean chunks retrieved per episode). CQL/BC retrieve broadly across all procedures; DPO retrieves selectively while maintaining CQL-level accuracy; IQL retrieves sparsely, focusing on high-relevance chunks.}
\label{fig:heatmap}
\end{figure}

\section{Discussion \& Conclusion}

Three distinct operating modes emerge from the same MDP formulation. CQL maximizes accuracy through exhaustive retrieval (92\%, 20 steps). IQL maximizes cost-efficiency through learned stopping (62.5\%, 3.4 steps). DPO bridges these two regimes, achieving CQL-level accuracy (92\%) with moderate retrieval depth (10.6 steps), yielding the best accuracy-efficiency tradeoff. The CQL--IQL divergence has an algorithmic explanation: CQL's conservative penalty pushes down Q-values for out-of-distribution actions, and the stop action ($a = K$) appears only once per episode in the offline data while retrieval actions appear multiple times, so CQL suppresses stopping as out-of-distribution. IQL's advantage-weighted policy extraction ($\beta = 10.0$) greedily selects the highest-return behavior from the dataset, and under $\lambda = 0.1$ with symmetric $\pm 1$ correctness rewards, stopping early at 62.5\% accuracy yields higher expected return than exhaustive retrieval at 92\%. Reducing $\lambda$ to 0.05 partially mitigates this: IQL trained on the $\lambda = 0.05$ buffer achieves 82\% accuracy in 9.0 steps, but still falls short of DPO's 92\% accuracy in 10.6 steps, confirming that the preference-based approach addresses this tradeoff more effectively than reward tuning alone. Which mode is preferable depends on the deployment context: high-stakes decisions may warrant CQL's thoroughness, high-volume settings benefit from IQL's efficiency, and DPO offers a practical middle ground where both accuracy and cost matter.

The BC-CQL equivalence confirms that CQL's conservative penalty reinforces exhaustive retrieval present in the data rather than learning new strategies. DPO avoids this by operating in policy space, learning when to stop naturally. A key methodological finding is that adapting DPO to sequential MDPs requires transition-level rather than trajectory-level optimization (Section~\ref{sec:dpo}), paralleling the difference between sequence-level and token-level objectives in language modeling. Per-procedure analysis (Section~\ref{sec:per_proc}) reveals that nearly all errors concentrate on three imaging procedures with high cross-procedure semantic interference from shared billing terminology.

\textbf{Limitations.} Our corpus uses synthetic CMS-derived requests and a rule-based oracle, which may not capture the full complexity of real PA decisions. The offline dataset is collected from relatively simple behavior policies, and richer behavior data may yield better offline RL performance. DPO's transition-level formulation assumes that good per-step actions in winning episodes are independently preferable, which may not hold in all MDPs.

\textbf{Future work.} We plan to evaluate on production payer policies with real clinical data, explore curriculum-based data collection strategies to improve coverage of hard procedures, and investigate whether DPO's selective-accurate operating mode transfers to other retrieval-augmented decision-making domains. A promising extension is quality-aware step rewards of the form $r_t = -\lambda + \alpha \cdot \text{sim}(e_q, e_{a_t})$, where the agent is rewarded for retrieving high-relevance chunks rather than treating all retrievals equally.

\section*{Author Contributions}
\begin{itemize}
\item \textbf{Ruslan Sharifullin:} MDP formulation, CQL / IQL / DPO implementation, training pipeline, convergence analysis, lambda ablation, experimental evaluation, off-policy evaluation (WIS, FQE), statistical significance test, per-procedure analysis, poster, figure generation.
\item \textbf{Maxim Gorshkov:} data pipeline, PA simulator, corpus construction, oracle design, offline dataset generation, test suite.
\item \textbf{Hannah Clay:} related work, introduction, discussion, conclusion, abstract, poster design, evaluation harness and metrics.
\end{itemize}

\section*{Source Code}
Source code, evaluation data, and the offline dataset are available at \url{https://github.com/rl-team/rl-adaptive-policy-retrieval}.\\ \\\\

\bibliography{bibliography}
\bibliographystyle{icml2018}

\end{document}